\documentclass{ws-procs975x65}
\usepackage{amssymb}
\usepackage{amsmath}

\def\phiq{{\phi}_q}
\def\phix{{\phi}_x}

\def\Journal#1#2#3#4{{#1} {\bf #2}, #3 #4}

\def\etal{{\it et al.}}

\def\APJ{\em ApJ.}
\def\APP{\em Astropart. Phys.}

\def\PRL{\em Phys. Rev. Lett.}
\def\PRD{{\em Phys. Rev.} D}

\def\be{\begin{equation}}
\def\ee{\end{equation}}
\def\bea{\begin{eqnarray}}
\def\eea{\end{eqnarray}}

\begin{document}

\title{Solving Two Puzzles in One Go: Quintessence from a Decaying Dark 
Matter}

\author{Houri Ziaeepour}

\address{Mullard Space Science Laboratory\\
Holmbury, St. Mary, Dorking, \\ 
RH5 6NT, Surrey, UK\\ 
E-mail: hz@mssl.ucl.ac.uk}


\maketitle

\abstracts {The mystery of the Dark energy is not just its smallness. The 
extreme fine tuning of its density in the Early Universe such that it permits 
galaxy formation before creating a new inflationary epoch is its other 
unexplained aspect. Here we propose a model based on the very slow decay of 
a Super Heavy Dark Matter (SDM), presumed candidate for the origin of the 
Ultra High Energy Cosmic Rays (UHECRs). A small part of the remnants energy 
is in the form of a scalar field which 
decoheres and behaves like a quintessence field. This model does not need a 
special form for the potential and a simple ${\phi}^4$ theory or an axion 
like scalar is enough. The equation of state of the Dark Energy becomes very 
close to a cosmological constant and is very well consistent with 
observations.}

In {\it Classical} Quintessence models a number of important issues are left 
unexplained. The complex behaviour of the potential they need is difficult to 
obtain from the Standard Model of particle physics or its extensions. In its 
simplest form it can not provide the observed $w_q \lesssim -1$ and does not 
explain the extreme relation between the Dark Matter (DM) and the Dark 
Energy (DE) density in the Early Universe.

Most solutions which have 
been recently proposed for resolving the latter problem are focused on the 
interaction of the quintessence scalar field $\phiq$ with the dark matter ~\cite{interac3}~\cite{interac0}~\cite{massvar}~\cite{yukawa}. 
These models however don't have $w_q \lesssim -1$ (although the main idea can 
be applied to the more complex quintessence models with unconventional 
kinetic term or non-minimal coupling to gravity). Some lead to a variable 
mass for DM particles which is very strongly constrained by the 
CMB anisotropy 
observations~\cite{cmbdmdeint}.

What we propose here ~\cite {dmquin} is a model for dark energy somehow 
different from 
previous quintessence models. We assume that DE is the result of the 
condensation of a scalar field produced during very slow decay of a massive 
particle. The main motivation is the possibility of a 
top-down solution ~\cite {xpart}~\cite {wimpzilla} for the 
mystery of UHECRs~\cite {fly}~\cite{agasa}. If a very small part of the decay 
remnants which make the primaries of UHECRs is composed of a scalar 
field $\phiq$, its condensation can have all the characteristics of a 
quintessence field. Moreover, it has been demonstrated that the 
cosmological equation 
of state for a decaying dark matter in presence of a cosmological constant 
is similar to a quintessence with $w_q \lesssim -1$ ~\cite {snhouri}. The 
latest estimation of $w_q$ from high quality SN-Ia light curves is 
$w_q= -1.05^{+0.15}_{-0.20}$~\cite {highsn}. The mean value is exactly in 
the range 
predicted for a decaying dark matter with a lifetime $\tau \sim 5{\tau}_0$ 
where ${\tau}_0$ is the present age of the Universe ~\cite {snhouri} (The 
error bars however are too large to make a definitive conclusion possible). 
This lifetime for a $M_{dm} \sim 10^{24} eV$ can also explain the observed 
flux of UHECRs without violating the present limits on the high energy 
neutrinos or photons ~\cite {wimpzilla}. Therefore it seems that both 
observations point to a top-down solution which explains simultaneously the 
dark energy and the UHECRs. 

The effective Lagrangian can be written as:
\be
{\mathcal L} = \int d^4 x \sqrt{-g} \biggl [\frac {1}{2} g^{\mu\nu} 
{\partial}_{\mu} \phix {\partial}_{\nu} \phix + \frac {1}{2} g^{\mu\nu} 
{\partial}_{\mu} \phiq {\partial}_{\nu} \phiq - V (\phix, \phiq, J) 
\biggr ] + {\mathcal L}_J \label {lagrange}
\ee
where scalar fields $\phix$ and $\phiq$ are respectively SDM and the 
quintessence. The field $J$ presents collectively other fields. 
The term $V (\phix, \phiq, J)$ 
includes all interactions including self-interaction potential for 
$\phix$ and $\phiq$:
\be
V (\phix, \phiq, J) = V_q (\phiq) + V_x (\phix) + g {\phix}^2 {\phiq}^2 + 
W (\phix, \phiq, J) \label {potv}
\ee
The term $g {\phix}^2 {\phiq}^2$ is important because it is responsible for 
the annihilation of $X$ and back reaction of the quintessence field. 
$W (\phix, \phiq, J)$ presents interactions which contribute to the decay of 
$X$ to light fields and to $\phiq$ (in addition to what is shown explicitly 
in (\ref{potv})). After writing the dynamical equations for the fields, 
one obtains the following asymptotic solution for $\phiq$ when its time 
variation is slowed down:
\be
V_q (\phiq) = V_q (\phiq (t'_0)) + {\Gamma}_q {\rho}_x (t'_0) 
\int_{t'_0}^t dt \biggl (\frac {a (t'_0)}{a (t)}\biggr )^3 
e^{-\Gamma (t-t'_0)} 
\label {phislowsol}
\ee
where $\Gamma$ and ${\Gamma}_q$ are respectively the total decay width and 
the width for the decay of the SDM to $\phiq$. ${\rho}_x$ is the density of 
the dark matter and $t'_0$ is the initial time for this asymptotic regime. 
If the slow down happens long before the matter-radiation equilibrium, at 
$\sim 100 t'_0$ the energy density of $\phiq$ is $\sim 90\%$ of its final 
value.
\begin{figure}[ht]
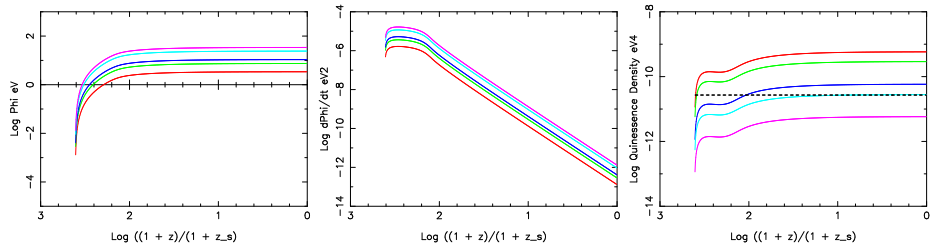

\centerline{\psfig{figure=quinphi.eps,angle=-90,width=4cm}
\psfig{figure=quindphi.eps,angle=-90,width=4cm}
\psfig{figure=quindens.eps,angle=-90,width=4cm}}
\caption{Evolution of quintessence field (left), its derivative (center) 
and its total energy density (right) for bottom to top curves 
${\Gamma}_0 \equiv {\Gamma}_q/\Gamma = 10^{-16}$ , $5 {\Gamma}_0$, 
$10 {\Gamma}_0$, $50 {\Gamma}_0$, $100 {\Gamma}_0$. Dash line is the 
observed value of the dark energy. $m_q = 10^{-6} eV$, 
$\lambda = 10^{-20}$.
\label {fig:quinevol}}
\end{figure}
This result has been also confirmed by the numerical solution of the 
evolution equations. Fig.\ref{fig:quinevol} and additional figures in 
\cite {dmquin} show that in a few orders of magnitude in redshift after the 
production of 
SDM, $\phiq$ approaches a saturation and its energy density changes very 
slowly. The details of the model depend on the self-coupling and mass, but 
general behaviour of the $\phiq$ field is quite stable. Parameters can be 
changed by many orders of magnitude without destroying the general behavior 
of the equation of state or the extreme relation between the energy density 
of dark energy and the total energy density in the early Universe. Therefore 
like other models with interaction between DM 
and DE, in this model the coincidence problem is solved without fine-tuning. 
Evidently this model does not explain the hierarchy of 
couplings and masses. But the relative value of these quantities are less 
extreme than the Cosmological Constant and the Plank mass. It has also been 
shown that the spatial perturbation of $\phiq$ is very small 
and decays with time. The effect on the CMB angular spectrum and Large 
Structures is very small and consistent with observations. All these 
properties are the consequence of the large lifetime of the SDM. 
In most quintessence models the scalar field is produced during 
inflation or reheating period in large amount and needs a negative 
exponential or a negative power function to 
control its contribution to the total energy of the Universe. 

The issue of decoherence of $\phiq$ in this model is not trivial and needs 
more investigation. The minimum condition for mode $k$ to decohere is 
~\cite {decohereinf0}: $k^2 / a^2 + {m_q}^2 \lesssim H^2$ 
where $H^2$ is the Hubble Constant. If the SDM exists and is produced during 
preheating just after the end of the inflation presumably at 
$T \sim 10^{14}eV-10^{16}eV$ scalars with mass $m \lesssim 10^{-6}eV$ can 
decohere. When the size of the Universe get larger, $\phiq$ stops 
decohering. This also helps having a very small dark energy density.

\end{document}